\documentclass[twocolumn]{aastex62}
\usepackage{amsmath,amstext}
\usepackage{units}
\usepackage{url}
\usepackage{xspace}
\usepackage{graphicx}

\newcommand{\E}[1]{\times10^{#1}}
\newcommand{\msol}{ \, M_\odot}
\newcommand{\smpy}{ \, M_\odot \, {\rm yr^{-1}}}

\newcommand{\bi}{\begin{itemize}}
\newcommand{\ei}{\end{itemize}}
\newcommand{\commentOut}[1]{}

\newcommand{\mesa}{\texttt{MESA}\xspace}

\shortauthors{Shen et al.}

\begin{document}


\title{\bf \Large{The Q Branch Cooling Anomaly Can Be Explained by Mergers of White Dwarfs and Subgiant Stars}}

\correspondingauthor{Ken~J.~Shen}
\email{kenshen@astro.berkeley.edu}

\author[0000-0002-9632-6106]{Ken~J.~Shen}
\affiliation{Department of Astronomy and Theoretical Astrophysics Center, University of California, Berkeley, CA 94720, USA}

\author[0000-0002-9632-1436]{Simon Blouin}
\affiliation{Department of Physics and Astronomy, University of Victoria, Victoria, BC V8W 2Y2, Canada}

\author[0000-0001-5228-6598]{Katelyn Breivik}
\affiliation{Center for Computational Astrophysics, Flatiron Institute, 162 Fifth Ave., New York, NY 10010, USA}


\begin{abstract}

\emph{Gaia}'s exquisite parallax measurements allowed for the discovery and characterization of the Q~branch in the Hertzsprung-Russell diagram, where massive C/O white dwarfs (WDs) pause their dimming due to energy released during crystallization.  Interestingly,  the fraction of old stars on the Q~branch is significantly higher than in the population of WDs that will become Q~branch stars or that were Q~branch stars in the past.  From this, Cheng et al.\ inferred that $\sim 6\%$ of WDs passing through the Q~branch experience a much longer cooling delay than that of standard crystallizing WDs.  Previous attempts to explain this cooling anomaly have invoked mechanisms involving super-solar initial metallicities.  In this paper, we describe a novel scenario in which a standard composition WD merges with a subgiant star.  The evolution of the resulting merger remnant leads to the creation of a large amount of $^{26}$Mg, which, along with  the existing $^{22}$Ne, undergoes a distillation process that can release enough energy to explain the Q~branch cooling problem without the need for atypical initial abundances.  The anomalously high number of old stars on the Q~branch may thus be evidence that mass transfer from subgiants to WDs leads to unstable mergers.

\end{abstract}



\section{Introduction}

As white dwarfs (WDs) evolve and cool, their interiors eventually undergo a liquid to solid phase transition, and they crystallize \citep{vanh68a}.  The process of crystallization releases latent heat and gravitational binding energy due to mixing induced by the phase separation between the oxygen-rich solid and  oxygen-poor liquid (see \citealt{baue23a} for a recent review).  These additional luminosity sources cause the cooling evolution of a WD to slow until the bulk of the WD has crystallized, which should result in a buildup of WDs on the cooling sequence  corresponding to the time of crystallization.

Using \emph{Gaia}'s revolutionary parallax measurements \citep{gaia16a,gaia18a}, \cite{gaia18b} and \cite{trem19a} resolved the WD crystallization overdensity in the Hertzsprung-Russell diagram, which was nicknamed the Q~branch due to its high concentration of DQ (carbon-polluted atmosphere) WDs.  However, several mysteries quickly became apparent.  For one, the hotter section of the Q~branch corresponds to the crystallization of ultramassive WDs with masses of $1.1-1.2 \msol$.  However, WDs arising from single-star evolution are composed of C/O up to $\simeq 1.05 \msol$ and O/Ne above this (\citealt{sies07a,lauf18a}; although see \citealt{alth21a}),\footnote{\cite{deni13b} proposed the formation of hybrid C/O/Ne WDs near this boundary due to the quenching of the inwardly propagating carbon-burning flame before it reaches the center of the WD, resulting in a C/O core surrounded by an O/Ne mantle.  However, \cite{leco16a}, \cite{broo17b}, and \cite{schw19a} pointed out several ways such a structure could not form and would not be stable even if it did form, so we do not consider hybrid C/O/Ne WDs further in this work.} and the massive end of the Q~branch coincides with the location of C/O WD crystallization, not O/Ne WD crystallization \citep{cami19a,baue20a,cami22a}.  Furthermore, \cite{chen19b} realized that the overdensity of WDs on the Q~branch is higher than predicted by the cooling delay due to the luminosity of latent heat release  and the overturn of oxygen during crystallization; another source of energy is required that yields an extra $\sim 8$-Gyr cooling delay for $\sim 6\%$ of the WDs passing through the Q~branch.

Several recent studies have proposed solutions to some of these puzzles \citep{blou20a,baue20a,capl20a,capl21a,blou21a,cami21a,wu22a,fleu22a,fleu23a}.  Of these, \cite{blou21a}'s work on $^{22}$Ne distillation is of particularly interest, but this process requires a very high $^{22}$Ne abundance to explain the multi-Gyr delay necessary to fully account for the properties of ultramassive WDs on the Q~branch.  In this work, we show that accurately accounting for compositional details during the merger of a C/O WD with the core of a subgiant star can explain the existence of ultramassive C/O WDs that undergo a long, multi-Gyr cooling delay during crystallization, driven by the distillation of $^{22}$Ne and $^{26}$Mg.


\section{Constraints on models to explain the cooling delay and a proposed solution}

The WDs that produce the overabundance on the Q~branch are undergoing C/O crystallization, because the process of O/Ne crystallization occurs at a higher core temperature and luminosity \citep{cami19a,baue20a}.  However,  single-star evolution does not produce C/O WDs as massive as the $\sim 1.2 \msol$ WDs at the blue end of the Q~branch.  One possible solution is the merger of two WDs that produces a $\sim 1.2 \msol$ merger remnant but avoids igniting carbon at any point, either explosively in a Type~Ia supernova during the merger or relatively quiescently at a later time as the merger remnant evolves \citep{it84,webb84,ni85,chen20b}.

During the merger, if no explosion occurs, the less massive WD is tidally disrupted into a rapidly rotating envelope surrounding the more massive WD.  Viscous stresses in the differentially rotating material convert kinetic energy into heat and expansion work, and the merger remnant relatively quickly becomes a degenerate core, consisting of the initially more massive WD, surrounded by a hot, spherical envelope made up of the less massive WD \citep{shen12,schw12}.  This envelope then radiates and contracts, its base temperature increases, and nuclear burning may be initiated at the interface between the envelope and the core.

\cite{schw21a} showed that any merger of two C/O WDs with a total mass $ \gtrsim 1.05 \msol$ leads to carbon ignition at the base of the contracting envelope, which eventually destroys the carbon throughout the core; thus, C/O+C/O WD mergers cannot be responsible for the  ultramassive C/O WDs on the Q~branch.  This leaves C/O + He WD mergers as a possibility, and indeed, as shown by \cite{wu22a}, carbon-burning can be avoided in the helium-burning ashes for C/O+He WD merger remnants  with masses up to $1.2 \msol$.

However, this progenitor scenario does not obviously explain the WDs with an extra-long cooling delay inferred by the kinematics and quantitative overabundance on the Q~branch.  The process of $^{22}$Ne distillation \citep{blou21a}, in which the exclusion of $^{22}$Ne from the solid phase eventually leads to a concentration of $^{22}$Ne in the interior of a crystallizing WD, presents an interesting possible solution, but the implied delay is still too short for standard compositions.  The overall $^{22}$Ne abundance needs to be several times higher than the value expected from helium-burning of CNO-processed material with solar abundances  ($X_{\rm 22Ne} \simeq 0.016$) to explain the extra delay with $^{22}$Ne distillation.  Thus, the question becomes: can the merger of a C/O and He WD result in the production of additional $^{22}$Ne during helium-burning?

The key to this question lies in the presence of relatively trace amounts of other elements besides the standard helium and $^{14}$N that make up the bulk of a He WD.  In particular, during and after the merging event, it is plausible that some  amount of carbon will be dredged up from the underlying core via Kelvin-Helmholtz instabilities and rotational mixing (see \citealt{hlw00} for a review of such mixing processes).  In addition, a small amount of hydrogen is present on the surface of He WDs and will be incorporated into the merger remnant's helium envelope.  As the merger remnant evolves and the envelope contracts and becomes hotter, the hydrogen will undergo CNO burning, which will process the dredged-up $^{12}$C primarily into $^{14}$N, since the proton capture onto $^{14}$N  is the slow step of the CNO cycle.  Once all of the hydrogen is consumed, the envelope contracts further and begins burning helium, converting the $^{14}$N into $^{22}$Ne; see \cite{clay07} and \cite{meno13a} for  analogous considerations of the compositions of R Coronae Borealis stars arising from lower-mass C/O + He WD merger remnants.

The resulting number abundance of $^{14}$N following the consumption of hydrogen is
\begin{align}
	n_{\rm 14N,f} &= n_{\rm 14N,i} + {\rm min} \left( \frac{n_{\rm H,i}}{2}, n_{\rm 12C,i} \right) ,
\end{align}
corresponding to a mass fraction of 
\begin{align}
	 X_{\rm 14N, f} &= X_{\rm 14N,i} + {\rm min} \left( 7 X_{\rm H,i},  \frac{ 7 X_{\rm 12C,i}  }{ 6 } \right) ,
\end{align}
where the subscripts $i$ and $f$ refer to the states before and after hydrogen-burning is complete, respectively.  The final $^{14}$N abundance depends on which element is the limiting factor, hence the min() function: if the initial hydrogen abundance is too low to convert all of the $^{12}$C into $^{14}$N, the excess $^{12}$C nuclei will undergo standard $\alpha$-captures during the subsequent helium-burning phase; conversely, if the initial carbon mass fraction is too low to absorb all of the hydrogen, the additional protons will use the existing $^{14}$N to go around the CNO cycle.

The solar abundance of CNO yields an initial $^{14}$N mass fraction in the helium envelope of $X_{\rm 14N,i} \simeq 0.01$.  We see that a hydrogen mass fraction of $X_{\rm H,i} > 0.001$ and a dredged-up carbon mass fraction of $X_{\rm 12C,i} > 0.01$ is sufficient to double the $^{14}$N mass fraction, consequently doubling the $^{22}$Ne in the helium-burning ash.

However,   a more careful consideration of the pre-merger binary evolution suggests that there is actually much less hydrogen available when the WDs merge.  At birth, a He WD with a mass of $0.2 \, (0.3) \msol$ possesses a hydrogen surface layer of $\simeq 8  \, (3)\E{-4} \msol$ \citep{istr16a}.  At face value, this could lead to a hydrogen mass fraction of $0.004 \, (0.001)$ in the helium-rich envelope of the merger remnant.  However, because of the low binary mass ratio and the non-degeneracy of the hydrogen-rich surface layers, the initial phase of hydrogen-rich mass transfer in these double WD systems is expected to be stable, leading to the transfer and subsequent ejection of the accreted hydrogen-rich layers in classical nova-like events.  It is only after the hydrogen has been transferred and ejected and  helium-rich accretion has begun that mass transfer is expected to become unstable during the subsequent helium novae, leading to a merger  \citep{shen13a,shen15a}.  Thus, the actual hydrogen mass fraction in the merger remnant's envelope is likely too low to lead to a significant boost in the $^{14}$N abundance.

Such He+C/O WD mergers will still occur and  yield ultramassive C/O WDs that crystallize along the Q~branch if the total merger remnant mass is $\lesssim 1.2 \msol$.  However, they will not have the necessary additional $^{22}$Ne to cause a long-enough cooling delay to  explain the observed Q~branch overdensity.  

Thus, we explore an alternative to this scenario, in the form of a C/O WD merging with a subgiant star crossing the Hertzsprung gap.  Such systems have been considered as supersoft X-ray sources and progenitors of Type Ia supernovae \citep{vand92,rapp94a,hach96a,yung96a,li97a,lang00a,hp04,tolo23a} because hydrogen-rich mass transfer in such a  system occurs on the thermal timescale of the donor, which places it close to or within the range for thermally stable nuclear burning on the surface of the accreting WD \citep{nomo82a,sb07,nomo07,wolf13a}.

Studies that find stable mass transfer and significant mass growth of the WD typically assume that mass transfer is fully conservative (e.g., \citealt{temm23a}) or that  mass transferred above the maximum rate for stable burning is blown away as an optically thick wind that is ejected with the specific angular momentum of the WD (e.g., \citealt{hach96a,hp04}).  However, the influence of the donor star on the formation of such a wind has not been considered.    As shown by \cite{shen22a} for the analogous case of hydrogen-rich classical novae, the gravitational influence of the companion precludes the formation of optically thick winds for most of the nova outburst, so that the binary's motion becomes the primary driver of mass loss in close binaries.  This results in a destabilizing loss of orbital angular momentum, which has strong observational support in the  cataclysmic variable mass distribution \citep{nele16a,schr16a}.  

Furthermore, some of the aforementioned studies completely neglect the influence of the helium novae that occur due to the buildup of helium ash from stable hydrogen burning.  Even those that do consider helium novae use helium accumulation efficiency calculations from \cite{kato99a,kato04a}, which again do not account for the influence of the companion on the formation of the helium nova wind.

\begin{figure}
  \centering
  \includegraphics[width=\columnwidth]{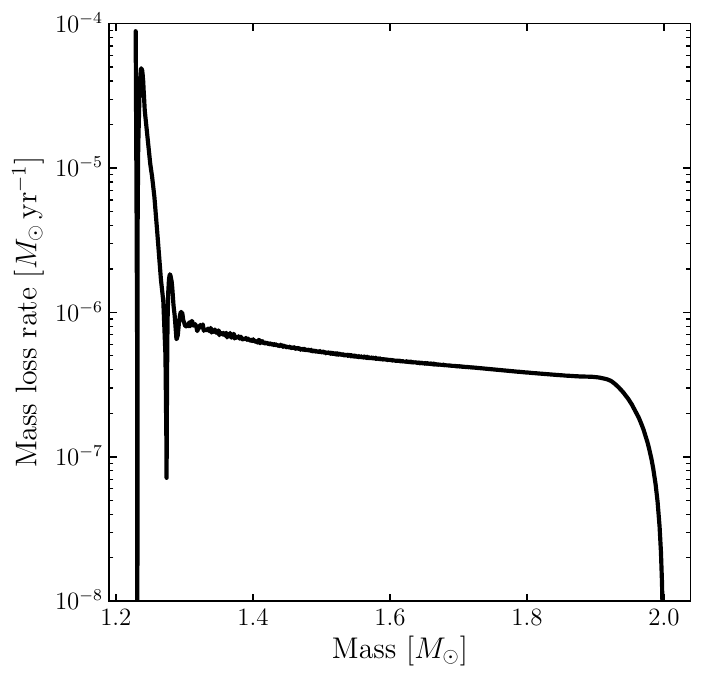}
	\caption{Mass loss rate vs.\ mass of an initially $2 \msol$ subgiant in a binary with a $1 \msol$ WD at an initial separation of $5 \, R_\odot$.  Mass transfer is assumed to be fully non-conservative, with 30\% of the angular momentum in the ejected mass coming from the binary orbit.}
	\label{fig:mdotvsm}
\end{figure}

When accounting for the influence of the donor, both of these processes -- mass transfer above the stable hydrogen-burning limit and helium novae -- result in losses of binary orbital energy and angular momentum that destabilize the system.  In Figure \ref{fig:mdotvsm}, we show the results of a binary stellar evolution calculation with \mesa\footnote{http://mesa.sourceforge.net, version 15140} \citep{paxt11,paxt13,paxt15a,paxt18a,paxt19a} of an initially $2 \msol$ subgiant with a $1 \msol$ WD companion (modeled as a point mass) at an initial separation of $5 \, R_\odot$.  As we are interested in the long-term evolution of the binary, spanning many hydrogen and/or helium novae, we assume the mass that is transferred from the subgiant is ultimately lost from the system, with the binary orbit providing 30\% of the necessary angular momentum to eject the material \citep{nele16a}.  We see that when the donor mass has decreased to $\sim 1.2 \msol$, the mass transfer rate oscillates dramatically and then diverges.  

\begin{figure}
  \centering
  \includegraphics[width=\columnwidth]{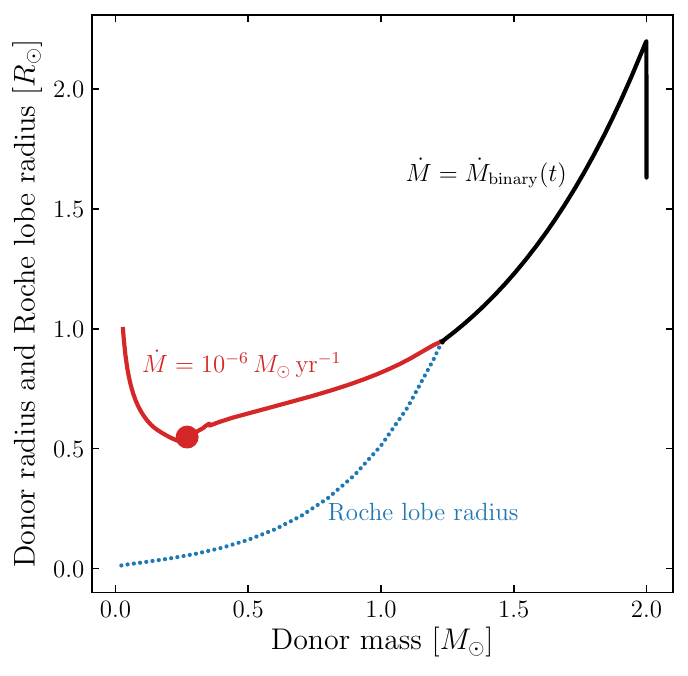}
	\caption{Radius of an initially $2 \, M_\odot$ subgiant  in a binary with a $1 \msol$ WD at an initial separation of $5 \, R_\odot$.  The initial radius evolution, shown as a black line, corresponds to the binary calculation shown in Figure \ref{fig:mdotvsm}.  After this phase, mass is removed at a constant rate of $10^{-6} \smpy$, resulting in the red line.  The blue dotted line shows the donor's Roche lobe radius assuming all of the angular momentum required for ejection comes from the binary orbit.}
	\label{fig:rvsm}
\end{figure}

In Figure \ref{fig:rvsm}, we show the donor's radius as a function of its mass.  The black line, labeled ``$\dot{M}=\dot{M}_{\rm binary}(t)$,'' shows the radius during the binary calculation described in the preceding paragraph and Figure \ref{fig:mdotvsm}.  We approximate the future evolution of the donor's radius, shown as a red line, by removing mass at a constant value of $10^{-6} \smpy$; while this mass loss rate is not quantitatively accurate, the corresponding mass loss timescale is shorter than the Kelvin-Helmholtz timescale of $\sim 10^8$\,yr and is thus qualitatively adequate for purposes of demonstration.  The point at which the helium-rich core is exposed is shown as a red circle.  The blue dotted line shows the donor's Roche radius assuming that all of the angular momentum required to eject the material comes from the binary orbit.

Notably, the subgiant donor's radius does not decrease significantly as mass is removed, and in fact increases once the non-degenerate helium-rich core is exposed.  At all times, the radius remains larger than the Roche radius.  This is in stark contrast to common envelope evolution involving a giant star: in that case, once the degenerate core is revealed, the radius decreases by over an order of magnitude and the donor may shrink within its Roche lobe, resulting in a detached close binary system.

Since the helium-rich core of the subgiant is not fully degenerate and is less condensed than the core of  a  giant, runaway mass transfer does not lead to the same outcome.  Instead, we expect the WD to inspiral all the way to the center of the subgiant and become the core of a single merger remnant with a helium-rich envelope polluted by hydrogen from the outer layers of the subgiant core and by carbon dredged up from the WD.

The situation is analogous to the predicted mergers of low-mass WD companions with main sequence donors due to dynamical friction induced by nova outbursts \citep{nele16a,schr16a,metz21a,shen22a}.  However, because of the more central concentration of the subgiant's helium-rich core and the subgiant's larger mass, we expect more material to be accreted by the WD than in the case of a main sequence star + WD merger, for which most of the main sequence star is likely ejected \citep{metz21a}.  We leave a quantitative calculation of the amount and composition of the accreted material to future multi-dimensional hydrodynamical simulations, but we conclude that it is at least plausible that the end result of mass transfer between a $ 2 \msol$ subgiant star and a $1 \msol$ WD is a merger between the WD and the core of the subgiant, and the ejection of the rest of the subgiant star.  Such a merger remnant would consist of a degenerate core (the former WD) surrounded by a hot, $\sim 0.2 \msol$ helium-rich envelope (the former subgiant core) polluted by hydrogen from the subgiant and carbon dredged up from the WD.


\section{The evolution of the remnant of a subgiant + white dwarf merger}
\label{sec:calc}

In this section, we describe the evolution of such a star formed by the merger of a $1.0 \msol$ C/O WD and the helium-rich core of a  subgiant.  We assume the  core is disrupted and becomes a hot $0.2 \msol$ envelope surrounding the WD, in an analogous way to the expected outcome of a double WD merger \citep{shen12,schw12}.  We again use \mesa, with the important addition of a 50-isotope nuclear network consisting of $n$, $^{1-2}$H, $^{3-4}$He,  $^7$Li, $^{7,9-10}$Be, $^8$B, $^{12-13}$C, $^{13-15}$N, $^{14-18}$O, $^{17-19}$F, $^{18-22}$Ne,  $^{21-25}$Na, $^{23-26}$Mg, $^{25-29}$Al, $^{27-30}$Si, $^{30-31}$P, and $^{31-32}$S.  Inlists and further details are described in the Appendix.

We first construct a $1.0 \msol$ C/O WD, starting with a $1.0 \msol$  helium core-burning star composed of 99\% $^4$He and  1\% $^{14}$N, by mass, as appropriate for the stripped core of a  solar-metallicity star that has undergone previous CNO-burning.  We then evolve the star, which first burns helium in the core and then in a shell as a helium giant.  Eventually, the helium shell becomes too low in mass to support helium-burning, and the star cools and becomes a WD.

\begin{figure}
  \centering
  \includegraphics[width=\columnwidth]{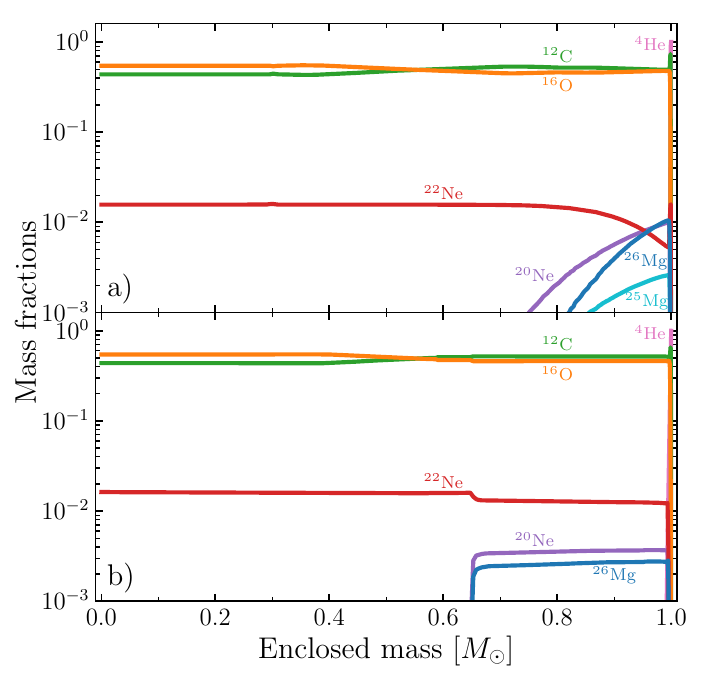}
	\caption{Mass fractions vs.\ enclosed mass within a $1.0 \msol$ C/O WD.  Panel~a) shows the composition when the  shell helium-burning luminosity has decreased to $1 \, L_\odot$.  Panel~b) shows profiles when the WD's  core temperature has cooled to  $10^7$\,K, $6\E{8}$\,yr after the time shown in panel~a).}
	\label{fig:1_0_abun}
\end{figure}

During the initial phases of helium-burning, the $^{14}$N captures $^4$He nuclei to become $^{18}$F, which $\beta$-decays to form $^{18}$O and then subsequently captures another alpha particle to become $^{22}$Ne.  Thus, the composition of the resulting helium-burning ash is a mixture of roughly $50/50$ C/O, by mass, plus a  trace amount of $^{22}$Ne with a mass fraction of $X_{\rm 22Ne} = 0.01 \times 22/14 = 0.016$.  However, as the core mass approaches $1.0 \msol$,  the helium-burning shell becomes hotter, and the $^{22}$Ne undergoes an additional alpha-capture to become $^{26}$Mg.  This can be seen in panel~a) of Figure \ref{fig:1_0_abun}, which shows the composition of the newly formed WD after the helium giant phase has ended and the helium-burning luminosity has decreased to $1 \, L_\odot$.  The $^{22}$Ne mass fraction in the helium-burning ash begins to decrease as the core mass grows, matching the $^{26}$Mg mass fraction after the C/O core reaches a mass of $0.96 \msol$.

Panel~b) of Figure \ref{fig:1_0_abun} shows the compositional profiles when the WD has cooled to a core temperature of $10^7$\,K, which occurs $6\E{8}$\,yr after the profiles shown in panel~a).  Notably, because of the relatively high mean molecular weight of the helium-burning ashes that contain $^{26}$Mg, the region underneath the shell is unstable to thermohaline mixing \citep{kipp80a}, which causes the outer $\sim 0.4 \msol$ underneath the helium layer  to homogenize.  However, as it will turn out, the quantitative details of the core's compositional profile at this time are not important because of future mixing.

To mimic the merger of a subgiant's core with the WD we have just constructed, we accrete $0.2 \msol$ of material with a composition of $X_{\rm 4He} = 0.95$ and $X_{\rm 14N} = 0.05$.  We thus assume that the previously discussed  conversion of extra accreted hydrogen (with a mass fraction $X \geq 0.006$) and dredged-up carbon (with a mass fraction $X_{\rm 12C} \geq 0.05$) into $^{14}$N occurs.  While it would have been preferable to include this phase of the evolution in our simulation, this proved too numerically difficult to calculate.  However, given the similar evolution calculated for lower-mass merger remnants \citep{clay07,meno13a}, which do not suffer from the extent of difficulties exhibited by higher-mass stars, we are  confident that such a calculation would result in the boosted $^{14}$N fraction that we use as our initial condition.

During the accretion phase, nuclear burning is disabled.  Once the desired mass is reached, we gradually relax this constraint until nuclear reactions have reached their proper values, at which point we begin the evolution of the merger remnant in earnest, and the star becomes a shell helium-burning giant.  Further details of the evolution and the inlists for the calculation are located in the Appendix.

\begin{figure}
  \centering
  \includegraphics[width=\columnwidth]{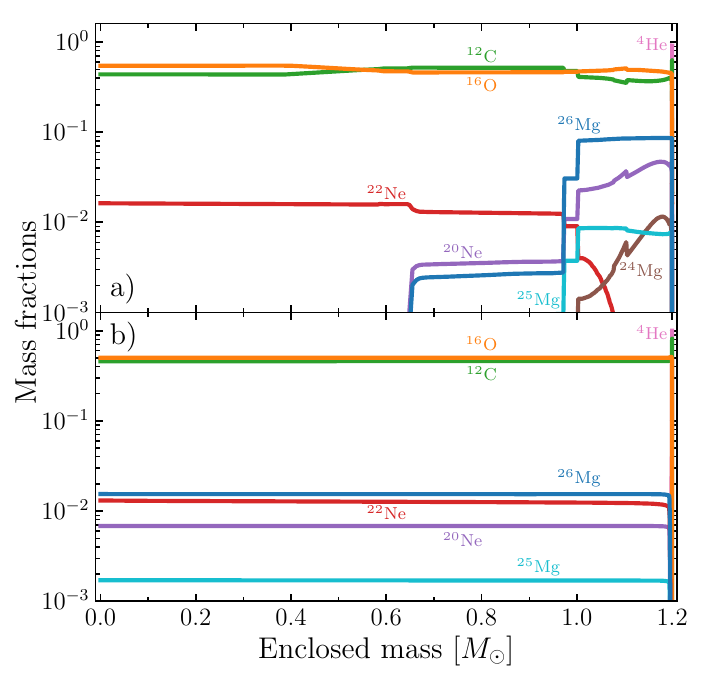}
	\caption{Mass fractions vs.\ enclosed mass within  a  merger remnant initially consisting of a $1.0 \msol$ C/O core with a $0.2 \msol$ envelope  composed of $95\%$ helium and $5\%$ $^{14}$N, by mass.  Panel~a) shows the composition when the  shell helium-burning luminosity has decreased to $1 \, L_\odot$.  Panel~b) shows profiles after the merger remnant has cooled to a core temperature of $1.5\E{7}$\,K, $5\E{8}$\,yr after the time shown in panel~a).}
	\label{fig:1_2_abun}
\end{figure}

Panel~a) of Figure \ref{fig:1_2_abun} shows the compositional profiles of the merger remnant after the shell helium-burning phase has ended and  the helium-burning luminosity has decreased to $1 \, L_\odot$.  As expected, the $^{14}$N in the shell, with an initial mass fraction of $X_{\rm 14N} = 0.05$, has been transformed into $^{26}$Mg, with a nearly constant mass fraction of $X_{\rm 26Mg} = 0.05 \times 26/14  = 0.09$.  We note that our nuclear network contains more massive elements than Mg, so these $^{26}$Mg nuclei could in principle capture $^4$He, but they do not do so under these conditions.

As before, the $^{26}$Mg in the helium-burning ashes leads to thermohaline mixing.  However, because of the larger mass of $^{26}$Mg-enriched ash and the  larger fraction of $^{26}$Mg within that ash, thermohaline mixing homogenizes the entire core, except for a small amount of helium that diffuses to the surface.  Panel~b) of Figure \ref{fig:1_2_abun} shows the compositional profiles $5\E{8}$\,yr  after the time shown in panel~a).  By this point, which is well before crystallization begins, the core has an essentially uniform composition. The  neutron-rich isotopes $^{22}$Ne and $^{26}$Mg have a combined mass fraction of $0.0125+0.0154=0.028$, 80\% more than if the $^{14}$N present in initially solar composition material were converted to $^{22}$Ne.  This number could be increased significantly if the amounts of accreted hydrogen and dredged-up carbon are  higher than we have assumed.  Future hydrodynamic simulations will help to better quantify these initial conditions.


\section{A cooling delay caused by $^{22}$Ne and $^{26}$Mg distillation}

\begin{figure}
    \includegraphics[width=\columnwidth]{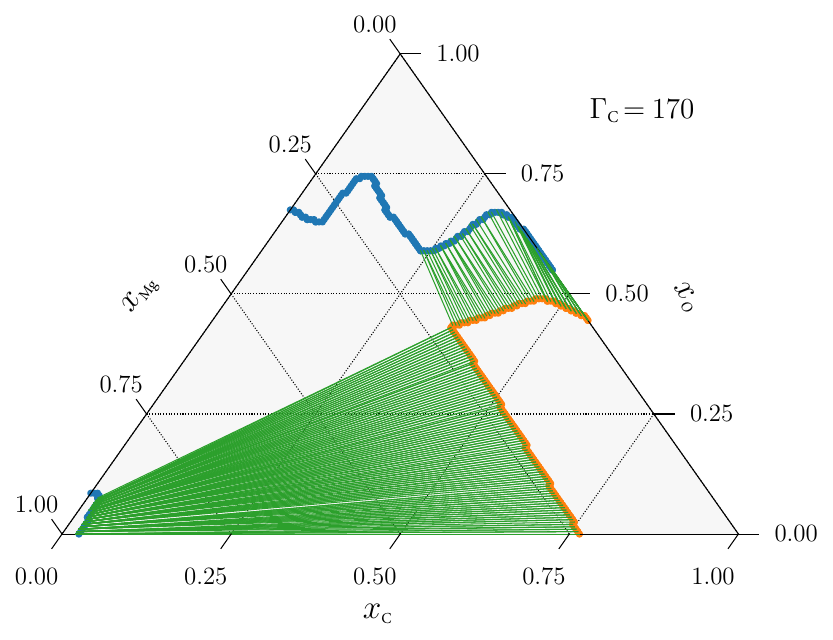}
    \caption{C/O/Mg phase diagram. The liquidus (orange) and solidus (blue) are connected by green lines that represent the composition difference between the coexisting liquid and solid phases. C/O mixtures with traces of magnesium are found close to the right axis. Note that number fractions are used in this figure.}
    \label{fig:COMg}
\end{figure}

Given its neutron-rich nature, $^{26}$Mg has the potential to initiate a distillation process similar to $^{22}$Ne distillation. To investigate this possibility, we calculate a C/O/Mg phase diagram using the semianalytic method detailed in \cite{medi10a} and \cite{capl18a}.\footnote{\url{https://github.com/andrewcumming/phase_diagram_3CP}} Figure~\ref{fig:COMg} shows a slice of that phase diagram for a plasma coupling parameter $\Gamma_{\rm C}$ close to the crystallization temperature of a mixture with equal masses of carbon and oxygen. The orange line represents the liquidus, the blue line corresponds to the solidus, and the green lines indicate how the composition changes upon freezing for different initial liquid compositions. For the low magnesium abundances relevant here ($x_{\rm Mg} \sim 1\%$), we observe that the solid phase is depleted in magnesium compared to the liquid: the green tie-lines are closer to the right axis at the solidus than at the liquidus. For a liquid with the initial $X_{\rm 26 Mg}=0.0154$ of our fiducial model, the solid is predicted to have a $^{26}$Mg mass fraction of only $\sim 0.003$. 

This $^{26}$Mg depletion is significant enough to render the solid phase buoyant, even after considering the oxygen enrichment in the solid. We verify this point by calculating the mass densities of the coexisting liquid and bcc solid phases for a  mixture with initially equal mass fractions of carbon and oxygen, $X_{\rm 26 Mg}=0.0154$, and $P = 4 \times 10^{25}\,{\rm erg}\,{\rm cm}^{-3}$, which corresponds to the pressure at the center of a $1.2\,M_{\odot}$ C/O WD. By employing the multi-component plasma free energy fits from \cite{dubi90a}, \cite{ogat93a}, and \cite{dewi03a} and accounting for the degenerate electron gas, we determine that the mass density of the solid is lower than that of the coexisting liquid by $0.08\%$.

When crystallization starts at the center of the star, these $^{26}$Mg-depleted crystals will float up and melt, displacing $^{26}$Mg-rich liquid inward in the process. This gradual transport of $^{26}$Mg towards the center liberates gravitational energy, just as in the case of $^{22}$Ne distillation \citep{iser91a,blou21a}. Based on the known C/O/Ne phase diagram, we also expect the floating solids to be depleted in $^{22}$Ne. Consequently, both $^{22}$Ne and $^{26}$Mg will distill, even if the initial $^{22}$Ne mass fraction is lower than the distillation threshold value established in \cite{blou21a}.

The ultimate outcome of this phase separation process remains uncertain due to the increasing abundances of $^{22}$Ne and $^{26}$Mg in the central layers, which gradually reduce the reliability of the ternary C/O/Ne and C/O/Mg phase diagrams. Phase diagrams including all four components would be needed, and ideally those calculations should be performed using a simulation-based approach \citep[e.g.,][]{blou21c}.\footnote{Indeed, we note that significant discrepancies between the semianalytic C/O/Ne phase diagram of \cite{capl20a} and the simulations of \cite{blou21a} arise at high Ne mass fractions.} Nonetheless, it seems plausible that most of the star's $^{22}$Ne and $^{26}$Mg is transported to the center. First, we already know that this is the expected outcome of simple $^{22}$Ne distillation. Second, for $^{26}$Mg, the C/O/Mg phase diagram of Figure~\ref{fig:COMg} has a similar shape to the O/Ne/Fe phase diagram presented in \cite{capl23a}, where the transport of Fe to the center was found to be possible through a combination of distillation and precipitation. 

Assuming that both $^{22}$Ne and $^{26}$Mg are indeed transported to the center of the star, we can now assess the impact of their combined distillation on WD cooling.  We can approximate this effect as the distillation of a mass fraction of $^{22}$Ne equivalent to $0.0125 + (10/12) \times 0.0154 = 0.025$, where the factor $10/12$ accounts for the slightly lower neutron excess per unit charge of $^{26}$Mg compared to $^{22}$Ne. Using the same approach as \cite{blou21a}\footnote{The chemical structure of the star was adjusted to reflect the expected stratification of the merger remnant. Note also that we assume that distillation proceeds until $X_{\rm 22Ne} + X_{\rm 26 Mg}=0.3$.}, we find that distillation in this star would emit $1.8 \times 10^{48}\,$erg, thereby introducing a cooling delay of $\sim 4.2\,$Gyr. 


\section{The space density of WD + subgiant merger remnants}

In this section, we  calculate the local space density of subgiant + C/O WD merger remnants in order to compare to the estimated overabundance of  Q~branch WDs.  We use the binary population synthesis code, \texttt{COSMIC}\footnote{https://cosmic-popsynth.github.io, version 3.4.8} \citep{brei20a}, which implements the single star and binary evolution prescriptions contained in \cite{hurl00a}  and \cite{hurl02a}, respectively.   We focus on mergers between C/O WDs and subgiants such that the total mass of the WD and the subgiant's helium-rich core is between $1.05-1.25 \msol$.

The population of subgiant + C/O WD mergers that produce Q~branch WDs is predominantly formed through a single formation channel in which the WD progenitor fills its Roche lobe on the asymptotic giant branch and initiates a common envelope evolution, which dramatically shrinks the binary, thus allowing the companion to fill its Roche lobe as a  subgiant $ <1$\,Gyr later. Because of this, we simulate several populations with varying assumptions for the common envelope ejection efficiency ranging from $\alpha=0.1-1.0$. We initialize each population assuming solar metallicity, a $50\%$ binary fraction, an initial stellar mass function following \cite{krou01a}, a uniform mass ratio distribution \citep{maze92a}, a uniform eccentricity distribution \citep{gell19a}, and a log-normal orbital period distribution \citep[e.g.,][]{ragh10a}.

We find between  $ 10^{-5} - 10^{-4} $ subgiant + C/O WD mergers  per unit solar mass of stars, depending on the common envelope ejection efficiency.  If we assume a constant star formation rate for the past 10\,Gyr \citep{cuka22a} and an extra 1\,Gyr of cooling to crystallization temperatures of the ultramassive C/O WDs that arise from these WD + subgiant mergers, we can calculate the number of C/O WDs with extra cooling delays of 4.2\,Gyr currently on the Q~branch for the $4\times 10^6 \, M_{\odot}$ of stars within 250 pc, a distance to which \emph{Gaia}'s sample of Q~branch WDs is essentially complete. 

We obtain between $20 - 200$ C/O WDs with extra cooling delays on the Q~branch at present; there are 125 such WDs if we assume  $\alpha=0.3$, as suggested by observational studies of  progenitors of binaries with at least one WD \cite[e.g.,][]{zoro10a,sche23a}.  These numbers are  comparable to the $\sim 240$ nearby Q~branch WDs with extra cooling delays estimated by \cite{chen19b}.  Given the uncertainties in our assumptions, the rough agreement between the number of observed Q~branch WDs and our binary population synthesis estimate suggests that mergers between subgiants and C/O WDs could produce a significant fraction, if not all,  of the Q~branch WDs with delayed cooling times.


\section{Conclusions}

In this Letter, we have proposed a novel progenitor channel to explain the overabundance of crystallizing, ultramassive ($\sim 1.2 \msol$) C/O WDs on the Q~branch: mergers of $\sim 1.0 \msol$ C/O WDs with the helium-rich cores of subgiant stars.  The resulting merger remnant consists of a degenerate C/O core surrounded by a hot, helium-rich envelope, with trace amounts of hydrogen from the outer layers of the subgiant's core and carbon dredged up from the WD.  These pollutants undergo CNO-burning and subsequently add to the existing $^{14}$N mass fraction.  When the envelope begins helium-burning, the $^{14}$N is burned past $^{22}$Ne to $^{26}$Mg.

After helium-burning is completed, the merger remnant cools into an ultramassive C/O-dominated WD.  The high mean molecular weight of the $^{26}$Mg-rich ash triggers thermohaline mixing, which homogenizes the core.  Finally, when crystallization begins, the relatively large amounts of $^{22}$Ne and $^{26}$Mg throughout the core lead to a distillation process that yields a $ \sim 4.2$-Gyr cooling delay, causing the pileup of ultramassive C/O WDs on the Q~branch.  The expected number of these merger remnants is in rough agreement with the estimated number of observed Q~branch WDs with extra cooling delays.

This progenitor scenario is particularly attractive because it does not require a super-solar initial metallicity: the large $^{14}$N abundance that eventually leads to the creation of $^{26}$Mg,  distillation, and the cooling delay is generated during the merger and subsequent thermonuclear evolution.  However, because the abundances of hydrogen and carbon that enrich the helium envelope following the merger are unknown, the exact amount of $^{26}$Mg produced and thus the length of the cooling delay are uncertain.  Better estimates of these values await future multi-dimensional hydrodynamic merger simulations.

Mass transfer from a subgiant star to a C/O WD has previously been assumed to increase the WD's mass to the Chandrasekhar limit, resulting in a Type Ia or Iax supernova.  If instead these systems merge unstably, they will be accompanied by much dimmer transient events and the eventual creation of atypically long-lived, ultramassive, crystallized WDs with $^{22}$Ne- and $^{26}$Mg-rich centers.  Thus, such systems may not die in explosions but instead may linger for billions of years as exotic probes of unstable binary mass transfer, rare thermonuclear reactions, and chemical mixing and distillation.


\software{\texttt{matplotlib} \citep{hunt07a}, \mesa \citep{paxt11,paxt13,paxt15a,paxt18a,paxt19a}, \texttt{COSMIC} \citep{brei20a}}

\begin{acknowledgments}

We thank Evan Bauer, Sihao Cheng, and Alison Miller for helpful discussions, the anonymous referee for useful comments, and the KITP for hosting the program ``White Dwarfs as Probes of the Evolution of Planets, Stars, the Milky Way and the Expanding Universe'' during which this research was initiated.  This work was supported by NASA through the Astrophysics Theory Program (80NSSC20K0544) and by the NSF under Grant No.\ PHY-1748958.  S.B.\ is a Banting Postdoctoral Fellow and a CITA National Fellow, supported by the Natural Sciences and Engineering Research Council of Canada (NSERC).  The Flatiron Institute is supported by the Simons Foundation.  This research used the Savio computational cluster resource provided by the Berkeley Research Computing program at the University of California, Berkeley (supported by the UC Berkeley Chancellor, Vice Chancellor of Research, and Office of the CIO).

\end{acknowledgments}


\appendix

In this Appendix, we describe the \mesa inlists used in this work to perform stellar evolution calculations.

The following inlist evolves a binary consisting of a $2 \msol$ subgiant and a $1 \msol$ WD (modeled as a point mass) at an initial separation of $5 \, R_\odot$.  Any mass transferred to the WD is assumed to be lost from the system via a combination of optically thick winds and hydrogen- and helium-powered novae.  The ejected mass is assumed to interact with the binary components, such that 30\% of the angular momentum necessary to eject the material comes from the binary's orbital angular momentum.

\

{\ttfamily
\noindent \&binary\_controls

\

   m1 = 2.0  ! subgiant mass in Msol

   m2 = 1.0 ! WD mass in Msol

   initial\_period\_in\_days = -1 ! set negative to use initial separation instead

   initial\_separation\_in\_Rsuns = 5

\

  ! assume all mass transferred to the WD is eventually lost from the system

    mass\_transfer\_beta = 1.0

\

  ! use other\_extra\_jdot routine  to subtract orbital angular momentum such that 30\% of the angular momentum necessary to eject the material comes from the orbit

   use\_other\_extra\_jdot = .true.

\

  ! set other angular momentum loss mechanisms to zero for simplicity

   do\_jdot\_gr = .false.

   do\_jdot\_mb = .false.

\

\noindent / ! end of binary\_controls namelist
}

\

In order to construct the C/O WD described in Section \ref{sec:calc}, we begin with a $1.0 \msol$ core helium-burning  star that then evolves into a helium shell-burning giant.  Once the helium shell decreases to a mass of $10^{-3} \msol$, oscillations in the outer $10^{-4} \msol$ cause the timestep to decrease significantly.  At this time, we set the opacity in the convective shell to a constant value of $0.2 {\rm \, cm^2 \, g^{-1}}$ to tame these oscillations and allow the star to evolve past this point.  This obviously changes the observational signatures of the star, but this is an acceptable tradeoff since we are only concerned with its compositional profile, which has essentially been set by this time.  Once the star has evolved and contracted to a radius of $0.1 \, R_\odot$, we allow the opacity in the outer regions to return to its normal values.  At this point, we also turn on diffusion, semiconvection, and thermohaline mixing and stop the evolution once the core temperature reaches $10^7$\,K.  The inlist for this calculation follows:

\

{\ttfamily
\noindent \&star\_job

\

      ! change the initial composition to a helium star; the composition is specified in \&controls

      relax\_initial\_to\_xaccrete = .true.

\

      ! change the nuclear reaction network to allow for expanded helium-burning

      change\_net = .true.

      new\_net\_name = '50\_iso.net'

\

      ! set the outer optical depth to be  10 x 2/3 to improve stability

      set\_to\_this\_tau\_factor = 10.0

      set\_tau\_factor = .true.

\

\noindent / ! end of star\_job namelist

\

\

\noindent \&controls

\

  ! disable gold tolerances to speed up the calculation

    use\_gold\_tolerances = .false.

\

      ! turn on MLT++ to improve stability

     okay\_to\_reduce\_gradT\_excess = .true.

\

   ! set initial mass

    initial\_mass = 1.0

\

  ! set initial composition as appropriate for a solar composition helium star

    accrete\_same\_as\_surface = .false.

    accrete\_given\_mass\_fractions = .true.

    num\_accretion\_species = 2

    accretion\_species\_id(1) = 'he4'

    accretion\_species\_xa(1) = 0.99

    accretion\_species\_id(2) = 'n14'

    accretion\_species\_xa(2) = 0.01

\

      ! when the helium shell decreases to 1e-3 Msol and the timestep becomes small, limit the maximum opacity

    ! opacity\_max = 0.2 

\

     ! when the photospheric radius decreases to 0.1 Rsol, remove the opacity limit and turn on mixing
     
      photosphere\_r\_lower\_limit = 0.1

    ! use\_Ledoux\_criterion = .true.

    ! alpha\_semiconvection = 0.1

    ! semiconvection\_option = 'Langer\_85'

    ! thermohaline\_coeff = 1.0

    ! do\_element\_diffusion = .true.

\

\noindent / ! end of controls namelist
}

\

To construct our $1.2 \msol$ merger remnant, we begin with our C/O WD and accrete $0.2 \msol$ of material with a composition of $X_{\rm 4He}=0.95$ and $X_{\rm 14N}=0.05$.    We disable nuclear burning during this phase.  After the desired mass has been accreted, we gradually turn nuclear burning back on while keeping the core at a constant temperature of $10^8$\,K.  This is hotter than the initial core temperature of $10^7$\,K, but such a cold temperature leads to very small thermal diffusion timesteps due to the steep temperature gradient.  However, since the helium-burning region and resulting ash are at an even higher temperature of $3\E{8}$\,K, the core temperature does not play a significant role in the evolution.

As before, oscillations in the convective envelope set in when the helium shell has decreased to a mass of $2\E{-4} \msol$, and we lower the opacity in the shell to $0.2 {\rm \, cm^2 \, g^{-1}}$.  The opacity is  returned to its normal values and mixing mechanisms are turned on when the merger remnant has contracted to a radius of $0.05 \, R_\odot$.  We use the default prescription for thermohaline mixing, but the choice is unimportant as the core's composition is homogenized long before crystallization occurs.  The inlist for the remnant evolution follows:

\

{\ttfamily
\noindent \&star\_job

\

      ! change the nuclear reaction network to allow for expanded helium-burning

      change\_net = .true.

      new\_net\_name = '50\_iso.net'

\

      ! set the outer optical depth to be  10 x 2/3 to improve stability

      set\_to\_this\_tau\_factor = 10.0

      set\_tau\_factor = .true.

\

\noindent / ! end of star\_job namelist

\

\

\noindent \&controls

\

  ! disable gold tolerances to speed up the calculation

    use\_gold\_tolerances = .false.

\

      ! turn on MLT++ to improve stability

     okay\_to\_reduce\_gradT\_excess = .true.

\

  ! accrete 14N-enriched helium shell until the merger remnant reaches 1.2 Msol

    mass\_change = 1d-6

    max\_star\_mass\_for\_gain = 1.2

    accrete\_same\_as\_surface = .false.

    accrete\_given\_mass\_fractions = .true.

    num\_accretion\_species = 2

    accretion\_species\_id(1) = 'he4'

    accretion\_species\_xa(1) = 0.95

    accretion\_species\_id(2) = 'n14'

    accretion\_species\_xa(2) = 0.05

\

  ! disable nuclear burning during accretion and relax it back to normal after reaching 1.2 Msol

    max\_abar\_for\_burning = -1

    ! eps\_nuc\_factor = 1d-15

    ! dxdt\_nuc\_factor = 1d-15

\

   ! keep the 1.0 Msol core at a constant temperature with other\_energy routine until nuclear burning has been returned to normal

    use\_other\_energy = .true.

\

  ! increase varcontrol\_target near the end of helium-burning, when the helium shell has decreased to 4e-3 Msol, to speed up the calculation

   !  varcontrol\_target = 1d-3

\

      ! when the helium shell decreases to 2e-4 Msol and the timestep becomes small, limit the maximum opacity and reset varcontrol\_target back to 1d-4

    ! opacity\_max = 0.2 

\

     ! when the photospheric radius decreases to 0.05 Rsol, remove the opacity limit and turn on mixing
     
      photosphere\_r\_lower\_limit = 0.05

    ! use\_Ledoux\_criterion = .true.

    ! alpha\_semiconvection = 0.1

    ! semiconvection\_option = 'Langer\_85'

    ! thermohaline\_coeff = 1.0

    ! do\_element\_diffusion = .true.

\

\noindent / ! end of controls namelist
}

\



\end{document}